# Heavy-Duty Electric Vehicles Contribution for Frequency Response in Power Systems with V2G

Xiaojie Tao
*Smart Grid Energy Research Center*
*University of California, Los Angeles, USA*
Los Angeles, USA
xiaojietao@g.ucla.edu

Yaoyu Fan
*Smart Grid Energy Research Center*
*University of California, Los Angeles, USA*
Los Angeles, USA
yaoyuf@g.ucla.edu

Zhaoyi Ye
*Smart Grid Energy Research Center*
*University of California, Los Angeles, USA*
Los Angeles, USA
zhy177@g.ucla.edu

Rajit Gadh
*Smart Grid Energy Research Center*
*University of California, Los Angeles, USA*
Los Angeles, USA
gadh@ucla.edu

*Abstract*— The integration of heavy-duty electric vehicles (EVs) with Vehicle-to-Grid (V2G) capability offers a promising solution to enhance grid stability by providing primary frequency response in power systems. This paper investigates the potential of heavy-duty EVs to support the California power grid under different charging strategies: immediate, delayed, and constant minimum power charging. Simulation results demonstrate that both V2G-capable EVs and non-V2G modes have great potential to provide primary frequency response, with V2G-capable EVs exhibiting especially strong contributions. The study highlights the influence of charging strategies, control modes, and grid conditions on EV contributions to grid stability, emphasizing their critical role in mitigating the adverse effects of renewable energy penetration.

*Keywords—Vehicle-to-Grid (V2G); Primary frequency response; Grid stability; Heavy-duty electric vehicles; Smart Grid; Smart charging*

## I. Introduction

As renewable energy sources such as wind and solar continue to expand, power systems face increased challenges in maintaining frequency stability. These renewable technologies lack physical inertia due to the absence of rotating mass, which traditionally helped dampen frequency deviations following generation disturbances. The resulting low-inertia grid conditions make power systems more vulnerable to frequency fluctuations and less resilient during sudden load or generation changes.

At the same time, heavy-duty electric vehicles (HDVs) are becoming a key component of the electrified transportation sector [1]. In California, regulations mandate aggressive deployment of zero-emission HDVs to meet decarbonization goals. This presents not only a substantial increase in electricity demand but also a valuable opportunity to use HDVs as distributed grid-support assets—especially if they are equipped with Vehicle-to-Grid (V2G) capability [2].

V2G-enabled HDVs can provide fast and localized power injection or withdrawal during grid disturbances. Their large battery capacities and bidirectional power converters enable rapid frequency response, which can complement the declining contribution from conventional synchronous generators. However, coordinating large-scale HDV participation requires careful control of charging behavior and scheduling. Without such coordination, unplanned charging may exacerbate demand peaks or lead to underutilization of their response capability.

Existing studies on frequency regulation using electric vehicles have largely focused on light-duty EVs or simplified fleet-level assumptions [3]. However, HDVs differ significantly in energy capacity, duty cycles, charging infrastructure requirements, and scheduling flexibility. These differences affect their availability and effectiveness in providing grid services. In addition, there has been limited analysis of how different HDV charging strategies—when combined with V1G (charging-only) or V2G (bidirectional) control modes—impact system-level frequency stability under low-inertia grid conditions [4].

To address these gaps, this study simulates the California power system with integrated HDV fleets under various control strategies. Specifically, we evaluate three representative charging strategies: (1) immediate charging, (2) delayed charging, and (3) constant minimum power charging. Each is analyzed under two control modes—V1G and V2G—to assess their impact on primary frequency response during grid disturbances. The simulation model incorporates realistic system inertia conditions, HDV charging loads, and a representative generation mix.

By comparing frequency response across scenarios, the study highlights the trade-offs between operational convenience, grid support effectiveness, and control complexity. The results demonstrate the significant potential of V2G-enabled HDVs to mitigate frequency instability and reinforce the importance of optimized charging strategies for resilient grid integration.

## II. Frequency Control

Frequency stability is a critical aspect of power system reliability. It reflects the balance between power generation and demand, with the grid frequency typically maintained at around 60 Hz in most systems. Even minor deviations can indicate imbalances and, if left unmanaged, can escalate into severe system instability or blackouts [5].

Fig. 1 illustrates a typical grid frequency transient following a loss of generation. Under normal conditions, the grid frequency experiences only minor oscillations. However, a sudden loss of generation results in a rapid frequency decline [6]. To counteract this, the primary frequency response is activated almost immediately, leveraging the kinetic energy stored in the rotating masses of conventional generators to stabilize frequency. This response typically lasts a few seconds and helps prevent further decline [7].

Approximately 30 seconds after the initial event, the secondary frequency response begins. This involves adjustments from automatic generation control systems to restore the grid frequency to its steady-state value of 60 Hz. Together, these two responses form the backbone of frequency control, ensuring that the system remains operational during disturbances.

The increasing penetration of renewable energy, which lacks the inherent rotational inertia of traditional power plants, has significantly weakened grid stability [8]. As a result, alternative methods to support frequency control are becoming essential. This need has created opportunities for innovative solutions, such as leveraging the capabilities of EVs to provide dynamic and scalable frequency response services [9].

## III. Heavy Duty Electric Vehicles

Heavy-duty electric vehicles (HDVs) have distinct features that make them ideal for supporting grid frequency stability. Frequency deviation, similar to voltage deviation, reflects the real-time balance between power generation and demand and arises from sudden mismatches—often due to generator trips or renewable variability [10]. If left uncorrected, such deviations can cascade into widespread instability and equipment damage.

With battery capacities between 200–600 kWh, HDVs offer substantial energy reserves for absorbing or injecting power during frequency events. Unlike light-duty EVs, HDVs can provide both Vehicle-to-Grid (V2G) services by discharging power and charging-only (V1G) services by rapidly shedding load. This dual-mode flexibility allows them to play a vital role in grid stabilization.

Equipped with fast-reacting power electronics, HDVs can respond almost instantly to disturbances, far quicker than traditional generators. This rapid reaction is critical in the initial seconds after a grid event, preventing further frequency drops and easing the burden on slower secondary control systems.

HDVs also enhance system resilience through spatial and temporal distribution. Deployed across diverse locations and operating schedules, they create a decentralized and dispatchable frequency response asset. This reduces dependency on fossil-fuel-based spinning reserves and supports decarbonization goals.

In summary, HDVs combine fast response, high energy capacity, and operational flexibility, making them well-suited for frequency support. To realize this potential, charging behavior must be strategically managed.

HDVs generally operate on fixed schedules, starting each shift at full charge. Recharging must complete before the next shift, typically during off-duty hours [11]. This study evaluates three real-world charging strategies that affect how HDVs interact with the grid: (1) immediate charging, (2) delayed charging, and (3) constant minimum power charging.

Immediate charging begins at full power (100 kW) as soon as the vehicle returns from service, typically around 16:00–23:00. While this ensures full readiness, it adds load during peak hours, increasing grid stress, voltage deviation, and electricity costs under Time-of-Use (TOU) pricing.

Delayed charging defers charging to late night or early morning (e.g., 23:00–6:00), aligning with lower grid demand. This reduces congestion and TOU costs, as shown by the shift in peak charging load in Fig. 2.

Constant minimum power charging starts immediately but at a lower, sustained rate (e.g., 50 kW for 14 hours from 16:00–6:00). This strategy distributes charging evenly over time.

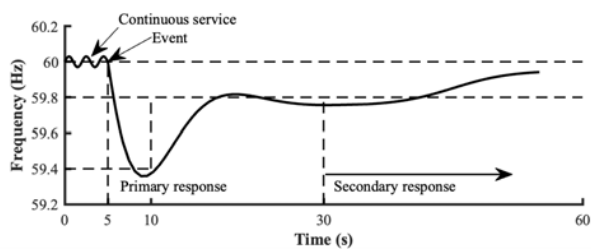

Fig. 1. Typical grid frequency response to a sudden generation loss, highlighting the primary and secondary frequency response phases. The primary response stabilizes the frequency dip caused by the event, while the secondary response restores the frequency to its nominal value of 60 Hz over time. This process demonstrates the critical need for rapid, dynamic support from sources such as Vehicle-to-Grid (V2G)-enabled heavy-duty electric vehicles during grid disturbances.

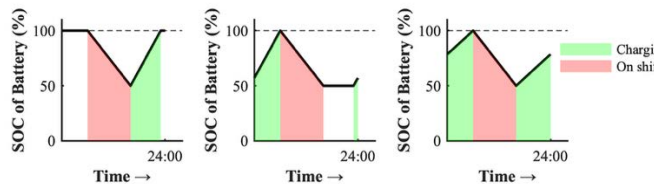

Fig. 2. State of Charge (SOC) profiles of heavy-duty electric vehicle (HDV) batteries under three charging strategies: immediate charging (left), delayed charging (middle), and constant minimum power charging (right).

TABLE I. CHARACTERISTICS OF THE THREE HDV CHARGING STRATEGIES.

| Charging strategy | Immediate | Delayed | Constant minimum power |
|---|---|---|---|
| Peak charging power (kW) | 100 | 100 | 50 |
| Charging duration (h) | 7 | 7 | 14 |
| Charging time | 16:00 – 23:00 | 23:00 – 6:00 | 16:00 – 6:00 |

TABLE II. SUMMARY OF CALIFORNIA'S GENERATION MIX AND INERTIA CONTRIBUTIONS DURING A SPECIFIC HOUR OF ANALYSIS. THE TABLE LISTS THE GENERATION SOURCES, THEIR INDIVIDUAL INERTIA CONSTANTS (HH), GENERATION SHARE, AND POWER OUTPUT.

| Generation source | H (s) | Generation share (%) | Generation power (MW) |
|---|---|---|---|
| Coal | 2.6 | 5.9 | 1,166 |
| Natural gas | 4.9 | 65.5 | 12,996 |
| Nuclear | 4.1 | 5.8 | 1,147 |
| Petroleum | 3.6 | 0.4 | 88 |
| Wind and solar | 0 | 4.0 | 809 |
| Hydro | 2.4 | 15.7 | 3,115 |
| Other | 0 | 2.6 | 509 |
| Effective H | 6.4 | 100 | 19,830 |

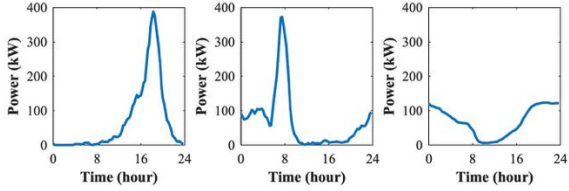

Fig. 3. Average power demand profiles of a heavy-duty electric vehicles (HDVs) fleet under three charging strategies: immediate charging (left), delayed charging (middle), and constant minimum power charging (right).

Table 1 compares the peak power, duration, and timing of these three strategies. While immediate and delayed charging both require 100 kW, the constant minimum strategy cuts peak power in half by doubling duration. This flatter demand profile is visualized in Fig. 3, which shows system-wide load curves under each strategy.

Each method presents trade-offs. Immediate charging maximizes readiness but stresses the grid. Delayed charging avoids congestion and saves cost but risks undercharging if delayed too long. Constant minimum power provides the most grid-friendly profile, though it may require longer planning and infrastructure support.

Ultimately, selecting the appropriate strategy depends on operational constraints, electricity pricing, and grid conditions. Strategic scheduling of HDV charging can balance fleet readiness with grid stability, enabling HDVs to function as valuable, responsive assets in a low-inertia grid.

## IV. CASE STUDY

To evaluate the potential of heavy-duty electric vehicles (HDVs) in providing primary frequency response, a case study was conducted using a Simulink-based model of the California power system shown in Fig. 4. The model incorporates key components, including synchronous power plants, system inertia, and HDV charging loads, represented by transfer functions. This approach enables a comprehensive analysis of the dynamic interactions between HDVs and the grid during frequency disturbances.

The system inertia is characterized by the effective inertia constant $H_{eff}$, defined as the weighted average of the individual inertia constants (H) of all generation sources, weighted by their power output [12]. Table 1 summarizes the inertia constants, generation shares, and power outputs for various energy sources, highlighting the reduced inertia contributions from renewable energy sources such as wind and solar, which lack the rotational mass of conventional generators. The effective inertia constant $H_{eff}$ is determined for a critical low-demand hour, February 28, 2021, at 20:00, when renewable energy penetration is highest and dispatchable generation is minimal.

HDVs can participate in frequency response using two distinct control modes:

- V1G Mode (Charging-Only): HDVs respond to a grid event by ceasing their charging, effectively shedding the instantaneous charging load.
- V2G Mode (Bidirectional): HDVs discharge power back into the grid during a grid event, contributing active power injection.

These modes enable HDVs to support grid stability under varying conditions [13]. The modeled system includes droop control for synchronous power plants, with parameters such as the droop coefficient R, generator time constant $T_G$, and turbine time constant $T_T$. Frequency deviations ($\Delta f$) are monitored by grid operators, triggering grid event signals for HDVs to activate their response mechanisms [14]. The frequency deviation Δf in this study refers to the single-step drop in system frequency triggered by a generation loss of 1,800 MW.

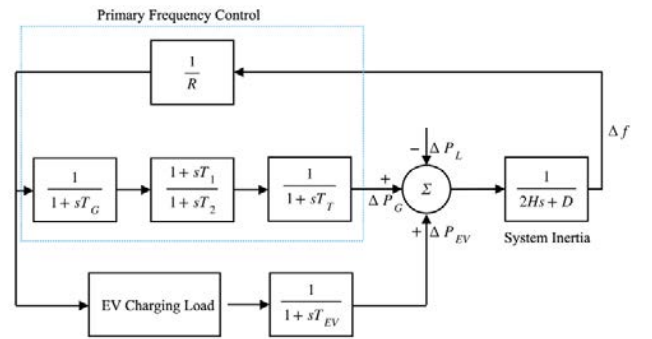

Fig. 4. Block diagram of the primary frequency control system incorporating heavy-duty electric vehicle (HDV) contributions.

## V. SIMULATION RESULTS

Simulations were conducted to assess the impact of HDVs on grid frequency stability under three charging strategies—immediate, delayed, and constant minimum power—and two control modes (V1G and V2G). The system was subjected to a simulated generation loss of 1,800 MW, with grid event signals sent when the frequency dropped below 59.7 Hz. The simulation model is implemented in Simulink and includes synchronous generators with droop control, HDV charging loads modeled as controllable elements, and a system inertia constant calculated from California's generation mix during a low-inertia hour. A step generation loss of 1,800 MW is introduced at t = 0, and grid event signals are triggered when the system frequency drops below 59.7 Hz. Frequency response under both V1G and V2G modes is evaluated across multiple HDV participation levels (20%–100%).

In immediate charging strategy, HDVs charged at full power immediately after their shifts create a concentrated load peak during grid peak hours. Fig. 5 shows the system frequency response under immediate charging. In V1G mode, increasing HDV participation raises the frequency nadir from 59.30 Hz to 59.55 Hz as more charging loads are shed. In contrast, V2G mode enables active power injection, resulting in significantly better performance—with nadirs reaching up to 59.75 Hz. The dynamic response is also faster, stabilizing frequency within 20 seconds. These results confirm that V2G provides not only deeper frequency support but also faster stabilization under disturbance.

Delayed charging, which shifts load to off-peak hours, performs similarly in frequency response but avoids grid congestion and reduces TOU costs (Fig. 6).

Constant minimum power charging delivers the most balanced outcome by spreading the load evenly. This strategy consistently maintains higher nadirs due to increased responsiveness and lower instantaneous demand (Fig. 7).

Figure 8 summarizes frequency nadir trends across the day. V2G mode consistently outperforms V1G, especially under high renewable penetration and low inertia conditions. Greater HDV participation leads to improved frequency resilience across all strategies.

The simulation results underscore the significant advantages of V2G mode compared to V1G mode in supporting grid frequency stability. By actively injecting power into the grid rather than merely halting charging, V2G mode enables HDVs to provide a more robust response during grid disturbances. This capability is particularly crucial during periods of low system inertia, where rapid and substantial power support is needed to mitigate frequency deviations. In contrast, while V1G mode contributes by shedding charging load, its impact is inherently more limited, highlighting the superior effectiveness of V2G technology in dynamic grid scenarios.

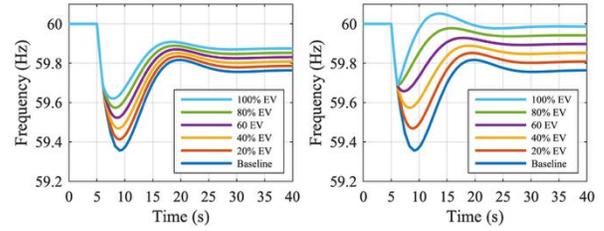

Fig. 5. Frequency response of the grid under the immediate charging strategy with varying levels of HDV participation in primary frequency control.

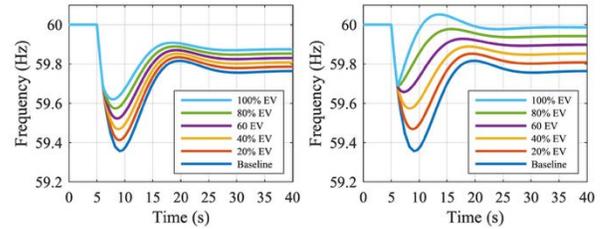

Fig. 6. Frequency response of the grid under the delayed charging strategy with varying levels of HDV participation in primary frequency control.

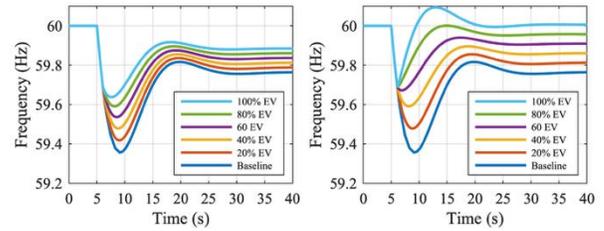

Fig. 7. Frequency response of the grid under the constant minimum power charging strategy with varying levels of HDV participation in primary frequency control.

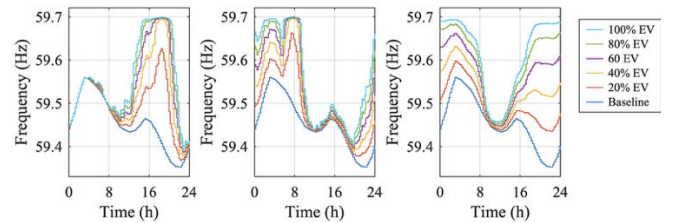

Fig. 8. Frequency nadir after a loss of generation every 15 minutes along a day.

The choice of charging strategy also plays a pivotal role in the overall grid impact and frequency response effectiveness. The constant minimum power charging strategy emerges as the most balanced approach, evenly distributing the charging load across off-peak hours, thereby reducing stress on the grid and enhancing frequency support during disturbances. The delayed charging strategy minimizes operational costs by avoiding peak demand periods, while immediate charging ensures operational readiness but imposes greater stress on the grid during peak hours. Additionally, the effectiveness of HDV responses is influenced by dynamic grid conditions, including varying demand patterns and renewable energy shares throughout the day. This variability underscores the necessity for adaptive and optimized charging strategies to fully leverage the potential of HDVs in enhancing grid stability.

## VI. Conclusions

This study evaluated the ability of heavy-duty electric vehicles (HDVs) to support primary frequency response in the California power grid under high renewable penetration. Simulations assessed multiple charging strategies and control modes using a low-inertia system model.

Results show that HDVs with Vehicle-to-Grid (V2G) capability significantly enhance frequency stability, outperforming V1G by providing active power support. Among the strategies, constant minimum power charging yields the most consistent performance by distributing load evenly and maintaining higher frequency nadirs.

These findings highlight the importance of optimized HDV charging coordination and V2G deployment in improving grid resilience under low-inertia conditions.

While simulation results demonstrate the technical feasibility of HDV-based frequency response, practical deployment requires coordinated scheduling, communication protocols, and regulatory alignment. Fleet operators must ensure vehicles are available and adequately charged before grid events, while utilities must provide incentives or compensation structures to encourage V2G participation. Furthermore, aligning HDV availability with high-risk grid periods (e.g., high renewable share and low inertia) remains a challenge. Future work should explore real-time coordination platforms and economic models to fully leverage HDVs as distributed grid assets.